\documentclass[runningheads]{llncs}
\usepackage[english]{babel}
\usepackage[T1]{fontenc}
\usepackage[utf8]{inputenc}
\usepackage{lmodern}

\usepackage{graphicx}
\graphicspath{{figs/}}
\usepackage{svg}
\usepackage{subcaption}
\usepackage{booktabs}
\usepackage{multirow}
\usepackage{makecell}
\usepackage{tikz}
\newcommand*\circled[1]{\tikz[baseline=(char.base)]{
    \node[shape=circle, draw, inner sep=1pt] (char) {#1};}}
\usetikzlibrary{arrows,positioning, decorations.pathreplacing}

\usepackage{textcomp}

\usepackage{color}
\usepackage[breaklinks]{hyperref}
\usepackage{comment}

\newcommand{\rfc}[1]{RFC #1~\cite{rfc#1}}

\begin{document}
\title{WHOIS Right? An Analysis of WHOIS and RDAP Consistency}
\author{Simon Fernandez \and Olivier Hureau \and Andrzej Duda \and Maciej Korczy\'nski}
\authorrunning{S. Fernandez et al.}
%\institute{Univ. Grenoble Alpes, Grenoble INP, LIG \\
\institute{Univ. Grenoble Alpes, CNRS, Grenoble INP, LIG, Grenoble, France \\
\email{\{firstname.lastname\}@univ-grenoble-alpes.fr}
}
%\url{}

\maketitle

\begin{abstract}

Public registration information on domain names, such as the accredited registrar, the domain name expiration date, or the abuse
contact is crucial for many security tasks, from automated abuse notifications to botnet or phishing detection and
classification systems.
Various domain registration data is usually accessible through the WHOIS or RDAP 
protocols---\textit{a priori} they provide the same data but use distinct formats and communication protocols.
While WHOIS aims to provide human-readable data, RDAP uses a machine-readable format.
Therefore, deciding which protocol to use is generally considered a straightforward technical choice,
depending on the use case and the required automation and security level.
In this paper, we examine the core assumption that WHOIS and RDAP offer the same data and that users can query them
interchangeably.
By collecting, processing, and comparing 164 million WHOIS and RDAP records for a sample of 55 million domain names, we reveal 
that while the data obtained through WHOIS and RDAP is generally consistent, 7.6\%
of the observed domains still present inconsistent data on important fields like IANA ID,  creation date, or nameservers.
Such variances should receive careful consideration from security stakeholders reliant on the accuracy of these fields.
%considered with care.
%Our findings raise questions about the effectiveness of systems based on the accuracy of this data and the
%confidence of the community in the quality of the information provided by some registrars.

 % \and coherence}

\end{abstract}%

\section{Introduction}\label{sec:intro}

%Various forms of malicious activities such as phishing scams, botnet operations, or malware distribution, involve domain names. Investigating these activities, or mitigating their impact 
%, or designing security metrics~\cite{korczynskiCybercrimeSunriseStatistical2018a}, 
%often require specific information regarding domain registrations. 
Malicious activities such as phishing scams, botnet operations, or malware distribution often involve the use of domain names. To investigate these activities and mitigate their impact, it is crucial to have access to specific information about domain registration.
%AD: strange to have a reference only to one attribut...
%MK: removed for now (the status of the domain~\cite{rfc8056})
%This information includes the creation date of a domain, the name registrant, the status of the domain, the expiration date, and the email addresses to be used in case of domain name abuse, etc. even though some personal information may be hidden to comply with the European General Data Protection Regulation (GDPR).
Essential information for investigating malicious activities related to domain names encompasses details such as the domain creation date, the registrant name, the sponsoring registrar, the domain status, the expiration date, email addresses designated for reporting domain name abuse, and other relevant data. However, in compliance with the European General Data Protection Regulation (GDPR)~\cite{gdpr} and the Temporary Specification of the Internet Corporation for Assigned Names and Numbers (ICANN) for generic Top-Level Domain (gTLD) registration data~\cite{icannTemporaryGTLD}, personal information pertaining to registrants is typically obscured or hidden.

Different entities involved in the domain registration process typically provide registration information through two protocols: WHOIS~\cite{rfc3912} and RDAP (Registration Data Access Protocol)~\cite{rfc9083}.
%
%WHOIS~\cite{rfc3912} provides access to registration and management information for domains, like the name of the
%owner (called registrant), the creation date, or the email addresses to contact if illegal activities are 
%detected on the domain.
%
%AD: j'ai mis cette partie dans background - ici il faut dire quelle est le probleme avec Whois/RDAP, pourquoi c'est important, et pourquoi ce n'est pas resolu.
%
%Even if there are historical reasons for the co-existence of two different protocols with two different formats theoretically providing access to the same data, almost all studies relying on the data~\cite{lauingerWHOISLostTranslation2016, gananWHOISSunsetPrimer2021, luWHOISWHOWASLargeScale2021,  maassEffectiveNotificationCampaigns2021} raise concerns regarding the protocols. 
Despite the historical reasons for the co-existence of two protocols, each having its own specific format, and theoretically providing access to the same data, numerous studies~\cite{lauingerWHOISLostTranslation2016,gananWHOISSunsetPrimer2021,luWHOISWHOWASLargeScale2021,maassEffectiveNotificationCampaigns2021} raised valid concern about the effectiveness and drawbacks of both protocols.

%Both protocols were designed to provide the same kind of information but there is no formal requirements that different data sources yield the same results.  Actually, the registration data may vary between registries and registrars, or between WHOIS and RDAP responses.
While both protocols were designed to provide registration information, there are no formal requirements mandating consistent results across different data sources. In practice, the registration data may vary between TLD registries, and registrars, as well as between the responses obtained from WHOIS and RDAP. This variability introduces an element of unpredictability with respect to the consistency and accuracy of the provided information.
%Moreover, most studies that use the registration data prefer one protocol over the other based on the ease of gathering and processing the data (query speed, parsing, the presence of an entry for each domain, etc.) without explaining the reasons for their choice. 

Furthermore, studies that use registration data tend to favor one protocol over the other without providing explicit justification, and they base their preference on factors such as data retrieval speed, parsing capabilities, the presence of WHOIS and RDAP records for each domain, and other convenience-related considerations.
Hence, an important issue emerges: to what degree do both protocols offer consistent information? Addressing this question requires a thorough and comprehensive analysis of how the data provided by the WHOIS and RDAP protocols align with each other.

%To the best of our knowledge, no previous work examined the underlying assumption that the information gathered from multiple sources is coherent.
To our knowledge, no previous research examined the assumption that information provided by WHOIS and RDAP is consistent.
%Still, many articles proposed classification algorithms, studied the domain behavior, or launched notification campaigns based on the 
%data from these protocols, thus depending on the truthfulness of the information they provide.
Nevertheless, many articles put forth classification algorithms, conducted studies on the domain behavior, or initiated abuse and vulnerability notification campaigns relying on  data obtained through these protocols. In doing so, they implicitly depend on the accuracy and consistency of the information provided by WHOIS and RDAP.

Our paper makes the following contributions:
\begin{itemize}
    %\item %In Section~????????????\ref{sec:background}, w
    %We overview the differences between WHOIS and RDAP and the reasons justifying
    %the presence of multiple servers and protocols providing the same data, highlighting how historical and technical decisions
    %led to a situation in which data consistency is not guaranteed.
    \item We provide an overview of the disparities between WHOIS and RDAP, shedding light on the rationale behind the coexistence of multiple servers and protocols for accessing registration data. Delving into the historical and technical aspects, we highlight the intricate choices that have led to the current state of uncertainty surrounding the assurance of data consistency. 
    \item 
    We undertake a comprehensive data collection encompassing WHOIS and RDAP records for more than 55 million domains. 
    Our focus is on parsing the fields commonly used in security and privacy studies.
    We will contribute all the collected registration data to the research community. 
    %We are dedicated to contributing all the collected registration data to the research community and a long-term provision of our API for WHOIS/RDAP 
    %collection and parsing, thus ensuring continuous access and support for researchers. 
    %We perform a wide data collection of WHOIS and RDAP entries for over 55 million domains and parse fields 
    %commonly used in security and privacy studies.  
    %We commit to contributing all the registration data to the research community and providing access to an API for WHOIS/RDAP collection and parsing.
    \item %We analyze the parsed fields, assess their consistency, and discuss potential sources of differences in content, raising awareness on how the community should be careful about trusting this data.
    We perform a thorough analysis of the parsed fields evaluating their consistency and deliberating over potential factors contributing to content variations. By doing so, we aim to raise awareness within the community about the importance of exercising caution with trust in registration data as 7.6\% of the observed domains presented inconsistencies in fields used by security and privacy studies.
    \item %We then run an analysis on the Nameservers field, cross-checking the collected data with results from the Domain Name System (DNS) to find which data source is more likely to provide accurate data.
    We conduct a comprehensive analysis of the nameservers field, cross referencing the gathered data with the results obtained from active DNS measurements. % of the Domain Name System (DNS). 
    Our aim is to determine which data source, whether WHOIS or RDAP, is more likely to provide accurate and trustful information.
    %\item We review how past and recent studies on security and privacy used the data collected in WHOIS and RDAP fields without validating its validity or consistency. We also discuss the attention that future work should pay when collecting, parsing, and using data from WHOIS and RDAP.
\end{itemize}

\section{Background}\label{sec:background}

We begin by providing background information on the administration of domain names and the collaborative processes within the DNS ecosystem.
Delving into the history of WHOIS and RDAP, we explore the reasons for their coexistence. Furthermore, we explain how to access registration data through both protocols, providing a clear outline of their respective procedures. Lastly, we elaborate on diverse approaches and challenges related to parsing WHOIS and RDAP.

\subsection{The Ecosystem of Domain Management and Registration}

The administration of a domain name entails the collaboration of multiple actors who collectively ensure the provision of all the necessary technical and administrative records vital for its operational use.
At the top of the Domain Name System (DNS), the Internet Assigned Numbers Authority (IANA) manages the 
root nameservers and delegates the management of each top-level domain (TLD) to different registries. 
%Country-code TLDs (ccTLDs) are managed by country-specific organizations like \texttt{.uk}, or \texttt{.fr} registry operators. 
Country-code top-level domains (ccTLDs) such as \texttt{.uk} and \texttt{.fr} are managed by country-specific organizations (registries) like Nominet (for \texttt{.uk}) or AFNIC (for \texttt{.fr}).
%while any organisation can manage generic TLDs (gTLDs).
In contrast, generic top-level domains (gTLDs) such as \texttt{.com} and \texttt{.business} can be managed by any organization that meets the necessary requirements~\cite{icannAgreement} and obtains authorization from the Internet Corporation for Assigned Names and Numbers (ICANN), like VeriSign Inc. (for \texttt{.com}) or Identity Digital (for \texttt{.business}).
%Registries manage their TLD zones and are responsible for the creation of new domains under their TLD. 
Registries are responsible for managing their top-level domain  zones and have the authority to create new domains under their TLD.
Each registry delegates the task of registering new domains to registrars, responsible for selling domains to users, referred to as registrants. 
When contacted by users, registrars collect and centralize user information, and communicate with the registry.
In the interaction between registrars and registries, a variety of protocols may be used with the Extensible Provisioning Protocol (EPP) \cite{rfc8056} commonly used for seamless communication.
%and contact the registry and communicate with the registry using the Extensible Provisioning Protocol (EPP) \cite{rfc8056}. 
%AD: develop EPP? 
%using the EPP protocol, which then writes the 
%different entries creating the domain.
The registry then generates the required records such as DNS ones and administrative details to create the domain.
For gTLDs under the ICANN agreement~\cite{icannAgreement} and the majority of ccTLDs, both the registry and the registrar make the registration information available to the public. 
%through the WHOIS and/or RDAP protocols.
This information is typically accessible through the WHOIS and/or RDAP protocols.
%The domain becomes usable when the different DNS records are added, but most registries and registrars also make additional information on the domain and the registration accessible through WHOIS and RDAP.

\subsection{Why Two Different Systems?}\label{subsec:history}
The existing WHOIS protocol as defined in \rfc{3912} published in 2004 formalized a practice in use since 1982~\cite{icannHistory}.
%It specified how a server could provide information on
%Internet entities (users, servers, domains, IPs, etc.), with a simple query/response protocol, but already pointed
%out that: ``\emph{WHOIS lacks many of the protocol design attributes, for example, internationalization and strong security,
%that would be expected from any recently-designed IETF protocol. This document does not attempt to rectify any of
%those shortcomings.}'', and only require the content to be ``\emph{in a human-readable format}''.
RFC 3912 established the guidelines on how a server could offer the information about various Internet entities, including users, servers, domains, and IP addresses with a straightforward query/response protocol. However, it recognized that the WHOIS protocol had certain deficiencies in terms of crucial design goals like internationalization and robust security, typically expected of IETF protocols.
RFC 3912 explicitly stated that it did not address these shortcomings and only required the content to be presented in a human-readable format.
The decision to retain the original design flaws in the WHOIS protocol can be attributed to historical reasons.
%The reasons for this choice are historical: the previous WHOIS system used since the early 70s, was already deployed on many servers and the IETF decided to go along with the original design flaws to avoid breaking backward  compatibility and forcing all operators to change their habits and systems, taking the risk of nobody using the new protocol (like what happened to the SPF DNS record\footnote{See \rfc{7208} Section 3.1}).
The original WHOIS system in use since the early 80s %70s, 
was already implemented on numerous servers. To maintain backward compatibility and prevent disruption to existing systems and practices, the IETF chose to accept the original design flaws rather than mandating widespread changes. This approach aimed to mitigate the risk of a new protocol facing low adoption rates, similar to what occurred with the \texttt{SPF} DNS record \cite{rfc7208}.
%
%After several years, the IETF started to work on a new protocol that could provide the same kind of information without having the same flaws as WHOIS, which resulted in the publication of \rfc{7482} that defined RDAP.

After several years, the IETF initiated efforts to design a new protocol aimed at providing domain registration information while addressing the limitations of WHOIS.
This endeavor culminated in 2015 with the publication of \rfc{7482} that specified RDAP. 
\rfc{7482}, along with subsequent extensions~\cite{rfc7483,rfc9082,rfc9083,rfc9224,rfc8982}, specifies the protocol emphasizing the provision of machine-readable data in the JSON format. It defines  %precise 
data types, keys, and encoding to ensure structured information.
Despite the introduction of RDAP, the WHOIS protocol has not been replaced, and both protocols continue to coexist, offering comparable data.

%The RFC, later extended by other documents~\cite{rfc7483, rfc9082, rfc9083, rfc9224, rfc8982}, 
%defines how the protocol should provide machine-readable data, represented as a JSON structure, with precise data types, 
%keys, and encoding.
%However, the WHOIS protocol was not replaced, and at the time of writing, both protocols still co-exist and advertise 
%the same kind of data.

\subsection{Data Access and Availability}\label{subsec:data_availability}

\rfc{3912} and \rfc{8521} define the WHOIS and RDAP data access protocols, respectively.  
The RDAP protocol operates over HTTP(s) using the REST paradigm and returns data in JSON format, 
while a WHOIS user needs to connect to a server over TCP on port 43 and receive a plain text response.

\begin{figure}[t]
    \centering
    \includegraphics[width=0.5\textwidth]{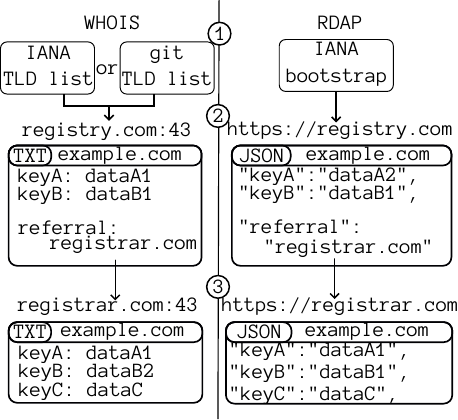}
    \caption{Referral system to obtain complete registration data}
    \label{fig:referral_graph}
\end{figure}

The registration data may be incomplete, and some registries may only offer minimal 
information---in this case, they are called ``thin'', in opposition to ``thick'' registries 
that directly provide the full registration data. 
This difference in the completeness of registration data remains valid for both WHOIS and RDAP. 
For instance, the \texttt{.com} registry provides minimal information and does not include 
the registrant organization data. 
To obtain complete information (with respect to GDPR), the user of both protocols may need to
follow referrals to one or several servers (see Figure \ref{fig:referral_graph}): 
they first need to locate the registry server (\circled{1}), then submit a query to the 
registry to obtain the registration information (\circled{2}), and optionally, retrieve more 
detailed data from the registrar~(\circled{3}).

For WHOIS queries, users can rely on command line tools provided by their system to bundle most
steps and referrals, like the Debian \emph{whois} package. 
On the contrary, there is no widely deployed command line tool to query RDAP databases.

% {\color{orange} A previous study~\cite{gananWHOISSunsetPrimer2021} highlighted how this referral chain and the type of protocol used to query 
% the servers could lead to significant differences in response times, with some servers taking up to a minute to provide
% the answer to the query.
% As a consequence, the ease of gathering WHOIS or RDAP data heavily depends on the number of referrals to follow, the quality 
% of  the third-party server lists providing the first server to query and the efficiency of each server involved in the process.}

The user needs to follow the steps below to retrieve registration information of 
\texttt{google.com} using RDAP:

\begin{itemize}
\item[\circled{1}] The user begins by retrieving the bootstrap configuration file from IANA,\footnote{\url{https://data.iana.org/rdap/dns.json}} as specified in RFC 9224. From this file, they obtain the URI of the \texttt{.com} RDAP server.
\item[\circled{2}] The user appends the string \texttt{domain/google.com} to the server URI obtained in step \circled{1}, and forms the query to retrieve the registry RDAP answer at \texttt{https://rdap.verisign.com/com/v1/domain/google.com}. (an illustration of the result can be found in Appendix~\ref{app:examples}, Figure~\ref{fig:rdap_example})
\item[\circled{3}] %Within the obtained JSON object, there is a link to the registrar's registration data. 
The returned JSON object contains a referral to the registrar server (in this example, MarkMonitor, Inc). The user can access this information at \texttt{https://rdap.mark monitor.com/rdap/domain/google.com}.
\end{itemize}

%By following this process, the user can retrieve the desired registration information for \texttt{google.com} using the RDAP protocol.

%As an example, to retrieve the registration information of  \texttt{google.com} using RDAP: The user retrieves the bootstrap configuration file from IANA\footnote{\url{https://data.iana.org/rdap/dns.json}} (as defined in RFC 9224) and retrieves the URI of the \texttt{.com} RDAP server (\circled{1}). He then appends the string \texttt{domain/google.com} to the server URI to retrieve the registry RDAP answer at \texttt{\url{https://rdap.verisign.com/com/v1/domain/google.com}} (\circled{2}). 
%A link to the registrar's registration data can be found in the JSON object: \texttt{\url{https://rdap.markmonitor.com/rdap/domain/google.com}} (\circled{3}).

For WHOIS, \rfc{3912} does not provide a bootstrap file for step \circled{1}. 
Instead, users can query the IANA WHOIS server at \texttt{whois.iana.org} to retrieve TLD-related 
information. The response includes the details about the TLD registry, in particular, 
the domain name of the WHOIS server for that zone.
As an example, let us examine the procedure involved in retrieving the registration information for the domain \texttt{google.com} using the WHOIS protocol:

\begin{itemize}
\item[\circled{1}] The user proceeds by querying the IANA WHOIS server for the \texttt{.com} TLD and locates the record \texttt{whois: whois.verisign-grs.com}. This information directs them to the VeriSign server.

\item[\circled{2}] Next, the user queries this server that provides registry WHOIS information for the domain \texttt{google.com} (the result is presented in Appendix~\ref{app:examples}, Figure~\ref{fig:whois_example}). 

\item[\circled{3}] Within this record, there is a referral to the registrar server \texttt{WHOIS Server: whois.markmonitor.com}.
The user can retrieve the most detailed registration data by querying this registrar WHOIS server.

\end{itemize}

Nevertheless, users may encounter problems when following this approach:
\begin{itemize}
    \item Certain WHOIS servers may require specific query flags. For example, the WHOIS server for the \texttt{.de} TLD expects the flags \texttt{``-T dn,ace''}.
    \item The IANA database may not always be up to date, resulting in inaccurate information about certain TLDs. For example, it does not provide a WHOIS server for the \texttt{.cm} TLD.
    \item In some cases, the TLD registry may not handle the registration information for domain names associated with public suffixes. For instance, the registry server \texttt{whois.nic.uk} for the \texttt{.uk} TLD does not manage the \texttt{.ac.uk} TLD,  managed instead by \texttt{whois.nic.ac.uk}.

\end{itemize}

For these reasons, the Debian \textit{whois} package\footnote{\url{https://tracker.debian.org/pkg/whois}} adopts a different approach. 
It uses a dedicated database that specifies servers responsible for the public suffixes and 
the corresponding flags to be used. The source code for this package is accessible in a 
collaborative GitHub repository.\footnote{\url{https://github.com/rfc1036/whois}}  While the 
repository allows anyone to propose modifications, it has been mainly maintained by Marco 
d'Itri since 1999.
This repository serves as a valuable alternative to the IANA WHOIS server, acting as a reliable starting point for retrieving WHOIS information (referred to as the \texttt{git TLD list} in Figure \ref{fig:referral_graph}, step~\circled{1}).

%For this reason, the Debian \textit{whois} package\footnote{\url{https://tracker.debian.org/pkg/whois}}  uses a different approach: it uses a dedicated database that indicates the server in charge of the public suffix and the appropriate flags to use. 
%
%Thanks to work done from Marco d'Itri done in 1999, bla bla bla i'm going to crous.
%The source code of the package is available in a GitHub repository\footnote{\url{https://github.com/rfc1036/whois}}. 
%This repository is collaborative and anyone can propose modifications. However, the package is maintained by only one person Marco d'Itri since 1999.

%We can see in Table ??????\ref{tab:tld_whois_rdap} 
%show the available whois and rdap server according to the different database  RDAP IANA website, 
%IANA's whois server and Git repositiry and presents those results in. We can observe 

We have retrieved the information from the RDAP bootstrap file, the GitHub 
repository %\footnote{\url{https://github.com/rfc1036/whois}} 
of the \textit{whois} package, and queried the server \texttt{whois.iana.org} for all active 
gTLD and ccTLD listed on the IANA website. %\footnote{\url{https://www.iana.org/domains/root/db}} We summarize those results in Table~??????\ref{tab:tld_whois_rdap}.
Table \ref{tab:tld_whois_rdap} shows that the GitHub repository provides 148 additional WHOIS 
servers compared to the IANA list. For instance, it includes a WHOIS server for the 
\texttt{.cm} TLD, not available on \texttt{whois.iana.org}.
The table also highlights the proportion of active gTLDs and ccTLDs that offer WHOIS and RDAP 
services. 
It is important to highlight that ccTLDs provide relatively less access to registration data than gTLDs.
In particular, the adoption of the RDAP protocol among ccTLDs is significantly low, 
accounting for only 9\%.
We can attribute the disparity between ccTLDs and gTLDs to the agreement established between 
gTLDs and ICANN~\cite{icannAgreement}. 
As per this agreement, registries have to offer access to registration data through the RDAP 
protocol. 
However, it does not require gTLDs to maintain WHOIS servers, and it does not apply to ccTLDs. 
Contrarily, the deployment of RDAP by ccTLD registries is influenced by various factors such 
as voluntary adoption, local regulations, and technical considerations.

\begin{table}[t]
    \centering
    \caption{Number of active TLDs providing RDAP and WHOIS servers}
    \vspace{+0.4cm}
    \begin{tabular}{c|c|c|c}
          &  RDAP & \multicolumn{2}{c}{WHOIS} \\
          Source & Boostrap & IANA & GitHub\\
         \hline
         ccTLD (309)  & 27 (9\%)  & 222 (72\%) & 231 (75\%)  \\ 
         gTLD (1152)  & 1152 (100\%) & 999 (86\%) & 1147 (99\%)  \\
    \end{tabular}
    \label{tab:tld_whois_rdap}
\end{table}

\subsection{Parsing Registration Data}\label{subsec:parse-data}
One of the primary motivations behind the design of RDAP is to address the 
inherent limitations of the WHOIS system, in particular, its vague and loosely defined 
``human-readable'' format for data. 
By incorporating the JSON-structured response format and well-defined data element features, 
among others, RDAP provides a more standardized, %structured,
 machine-readable approach to 
accessing registration data. This enhancement significantly improves the efficiency and 
reliability of parsing and extracting information from RDAP responses when compared to the 
traditional WHOIS system.

WHOIS data has %traditionally 
been presented in various formats, undergone frequent changes, and may even be expressed in  
the local language of the registrar or TLD registry (e.g., the Bolivian ccTLD \texttt{.bo} WHOIS  
records are written in Spanish).
The absence of normalization or implicit conventions raises a significant challenge when parsing WHOIS records, as highlighted in the studies that use WHOIS data~\cite{ghalebCyberThreatIntelligenceBased2022,maassEffectiveNotificationCampaigns2021,vissersParkingSensorsAnalyzing2015,liuWhoComLearning2015,luWHOISWHOWASLargeScale2021,lauingerWHOISLostTranslation2016}.

We can categorize traditional algorithms for parsing WHOIS data into two distinct approaches: 
templates and rules.
The template-based approaches, such as \texttt{Net::Whois}\footnote{\url{https://metacpan.org/pod/Net::Whois}} (Perl), \texttt{whoisrb}\footnote{\url{https://whoisrb.org/}} (Ruby), and \texttt{PHPWhois}\footnote{\url{https://github.com/SimpleUpdates/phpwhois}} (PHP), offer regular expression templates specifically tailored to each registry or registrar.
When using this approach, the user obtains WHOIS data from the registry, parses it using the relevant template for the TLD and registry, extracts any potential referral link to a registrar WHOIS server, and then retrieves and parses the registrar WHOIS data using the corresponding template. 
This approach is effective when the templates are available and regularly maintained. 
However, it becomes challenging when no template is available for a specific entity or 
if the format undergoes changes. Therefore, its success heavily relies on the quantity and 
quality of the templates, necessitating manual %crafting and frequent 
updates for each template.
%Due to their efficiency, template-based approaches are occasionally employed by research teams seeking specific fields from particular registrars. These teams create custom templates tailored to their specific requirements, allowing them to extract the desired data effectively.

Rule-based approaches such as \texttt{python-whois}\footnote{\url{https://pypi.org/project/python-whois/}} 
use a collection of predefined rules, regular expressions, and Natural 
Language Processing techniques to identify prevalent formats found in WHOIS records such as 
\texttt{Key: Value}, and extract as many fields as feasible. This approach is versatile and 
can be applied to any registrar without the need for dedicated templates. It may also 
accommodate format changes over time. However, it is generally less efficient compared to 
the use of custom-made templates~\cite{liuWhoComLearning2015}.

Previous work explored existing parsers to train machine-learning algorithms based on Natural 
Language Processing or used techniques like Conditional Random 
Field~\cite{mccallumEarlyResultsNamed2003} to automatically deduce the data structure and 
enhance the accuracy of field extraction. This approach demonstrated improved 
capabilities in extracting various fields from data.

\begin{figure}[t]
    \centering
    \includegraphics[width=0.9\textwidth]{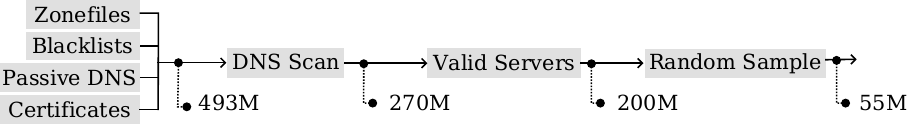}
    \caption{The stages of domain selection with the number of domains at each step}
    \label{fig:domain_selection}
\end{figure}

%AD: already said...
%Parsing registration data from WHOIS entries can be challenging due to the lack of normalization and implicit conventions. 
While the template-based and rule-based approaches offer some potential for obtaining registration data through WHOIS, they require regular maintenance and may be less efficient than RDAP. The introduction of RDAP offers a promising alternative for enhanced parsing efficiency and accuracy.

\section{Methodology}\label{sec:methodology}

In this section, we outline our methodology for collecting and parsing WHOIS and RDAP records. 
Considering the significant volume of data, we have meticulously designed our scheme to efficiently collect and parse registration data for a large number of domains within a reasonable time frame. All this is achieved while ensuring that WHOIS and RDAP servers experience minimal strain.
We begin by explaining the process of domain selection, as illustrated in Figure~\ref{fig:domain_selection}, followed by a comprehensive 
description of the WHOIS and RDAP parsing process. %techniques. 
Lastly, we provide an overview of how we have identified and analyzed discrepancies among the 
records.

\subsection{Domain Data Collection and Filtering
%Domain Selection
}\label{subsubsec:domain-selection}

%Collecting and parsing WHOIS and RDAP data is a painstaking process, using a lot of bandwidth and computation power,  because the scan of each domain involves several requests to servers, following the referrals and then parsing the collected results, with the difficulties described in Section~??????\ref{subsec:parse-data}. For these reasons, we limited ourselves to a subset of domains. The different steps and the number of domains at each stage are represented in Figure~??????\ref{fig:domain_selection}.

%\paragraph{List of active registered domains}
%First, we built a list of as many  %registered 
%domains as possible by aggregating several data sources:
\subsubsection{Compilation of registered domain names.}

First, we gathered an extensive list of domains by consolidating multiple data sources:
\begin{itemize}
\item gTLD zone files obtained from the ICANN Centralized Zone Data Service (CZDS),\footnote{\url{https://czds.icann.org}}

\item ccTLD zone files accessible via \texttt{AXFR} zone transfers (\texttt{.se}, \texttt{.nu}, \texttt{.li}, \texttt{.ch}),

\item Passive DNS feed from SIE Europe,\footnote{\url{http://sie-europe.net}}

\item Domain blacklists including SpamHaus,\footnote{\url{https://www.spamhaus.org}}
APWG,\footnote{\url{https://apwg.org}} OpenPhish,\footnote{\url{https://openphish.com}}
URLHaus,\footnote{\url{https://urlhaus.abuse.ch}} ThreatFox,\footnote{\url{https://threatfox.abuse.ch}} and SURBL,\footnote{\url{https://surbl.org}}

\item Google Certificate Transparency Logs,\footnote{\url{https://certstream.calidog.io}} which we continuously monitored to identify newly issued Transport Layer Security (TLS) certificates and extract the corresponding domain names.
    
\end{itemize}

%All these domains are aggregated and deduplicated, generating a list of 493M unique domain names. Because some sources can provide domains that may no longer be active (like passive DNS and blacklists), we run a DNS scan  on all domains, filtering out the ones without a valid \texttt{A}, \texttt{AAAA}, \texttt{NS} or \texttt{MX} entries, leaving 270M active domains.
All the collected domains are aggregated and deduplicated, resulting in a list of 493 million unique domain names. To guarantee the inclusion of only registered domains, we performed an active DNS scan on each domain, querying for \texttt{A} resource records using \texttt{zdns}~\cite{izhikevichZDNSFastDNS2022}, and exclude those for which the response is \texttt{NXDOMAIN} (non-existent domain).
%eliminating those lacking valid \texttt{A}, \texttt{AAAA}, \texttt{NS}, and \texttt{MX} entries—specifically, those returning \texttt{NXDOMAIN}. This process narrows down the list to 270 million registered domain names.

\subsubsection{Filtering domains with valid WHOIS and RDAP  servers.}
To study the inconsistencies between WHOIS and RDAP records, we carefully filtered out domains that lacked a recognized WHOIS or RDAP server. This filtering process involved cross-referencing the official IANA list~\cite{ianaTLD} and the GitHub repository, as detailed in Section \ref{subsec:data_availability}. After this filtering step, our dataset comprised 200 million domain names.

%Considering the scale of the measurement, s
Scanning all 200M domains would be a time-consuming process spanning several months, along with significant storage challenges. To address this, we opted to work with a representative subset of domains. 
%This subset was randomly selected from the pool of 200M domains, with the sample size determined to allow for the collection and parsing of %full 
%WHOIS and RDAP entries within a month.
This subset was randomly chosen from the pool of 200 million domains, with a sample size of 55 million domains carefully determined to facilitate the collection and parsing of WHOIS and RDAP records within a one-month time frame.

%By using the tools we developed and employed, we were able to process registration data for over 55M unique domain names. % in this sample. 
%To ensure the integrity of our findings and avoid any potential selection biases, we conducted additional checks to verify that the distribution of registrars and registries within the sample accurately reflected the broader domain population.

\subsection{Gathering and Parsing Resgistration Data}\label{subsec:processing-data}

\subsubsection{Data collection.}
%We collected the registration entries of all the selected domains.
%As outlined in Section~\ref{subsec:data_availability}, our initial step involved identifying the WHOIS and RDAP registry servers responsible for managing the top-level domain (TLD) of each domain.
%Once these servers identified, 
After identifying WHOIS and RDAP servers for the sampled domain names, we proceeded with the collection of the corresponding records.
We gathered the registration data of the selected domains between December 6th and December 31st, 2022.
During the collection process, we parsed each record to determine if it belonged to a ``thin'' registry that delegated a part of the data to a referral server, and follow the eventual referral. This step was iteratively repeated to ensure we obtained all versions of the registration data, following all referrals.
At the end, we successfully collected a total of 164 million unique records, covering information from over 55 million distinct domains.

%The collection of this many entries takes time, so w
To ensure accurate comparisons, %and minimize mismatches, 
we collected WHOIS and RDAP records of each domain within a narrow time window, typically under 1 minute. This prevents the comparison of records collected at different times and reduces discrepancies resulting from domain updates during the scanning process.
%
%As mentioned in Section~\ref{sec:background}, 
Moreover, some registrars impose query limits on IP addresses and enforce timeouts or blacklist IP addresses that exceed these limits. To ensure compliance and prevent any disruptions, we adjusted our data collection speed accordingly.

After the collection process, we carefully examined the gathered WHOIS and RDAP records. Any malformed responses (like invalid HTTP packets or JSON objects for RDAP) or timeouts were discarded, while valid responses underwent parsing for further analysis.

\begin{table*}
    \center
    \caption{Fields extracted from WHOIS and RDAP records}
    \vspace{+0.4cm}
    \begin{tabular}{r|c|c|c|c|p{2cm}}
        \multirow{2}{*}{Field} & \multirow{2}{*}{Data type} & \multicolumn{2}{c|}{Missing rate} & \multirow{2}{*}{Domain inconsistency} & \multirow{2}{*}{Used by} \\
        & & Records & Domains & & \\
        \hline
        \texttt{Nameservers} & Text & 3.2\% & 6.6\% & 573,790 (1\%) & \cite{christinDissectingOneClick2010,felegyhaziPotentialProactiveDomain2010,haoPREDATORProactiveRecognition2016} \\
        \texttt{IANA ID} & Integer & 5.9\% & 13.7\% & 106,813 (0.2\%) &\cite{affinitoDomainNameLifetimes2022,christinDissectingOneClick2010,lepochatPracticalApproachTaking2020b,duEverchangingLabyrinthLargescale2016} \\
        \texttt{Creation date} & Date & 0.8\% & 2.2\% & 3,138,024 (5.7\%) &\cite{ghalebCyberThreatIntelligenceBased2022,affinitoDomainNameLifetimes2022,lepochatPracticalApproachTaking2020b,duEverchangingLabyrinthLargescale2016} \\
        \texttt{Expiration date} & Date & 1.0\% & 2.7\% & 2,424,951 (4.4\%) & \cite{lauingerWHOISLostTranslation2016,lepochatPracticalApproachTaking2020b} \\
        % \texttt{Update date} & Date & 2,158,583 (4\%)  & Novel \\
        \multirow{2}{*}{\texttt{Emails}} & \multirow{2}{*}{Email} & \multirow{2}{*}{7.9\%} & \multirow{2}{*}{14.8\%} & \multirow{2}{*}{18,958,821 (34.5\%)} & \cite{duEverchangingLabyrinthLargescale2016,cetinUnderstandingRoleSender2016,luWHOISWHOWASLargeScale2021,vissersParkingSensorsAnalyzing2015} \\
        & & & & & \cite{ghalebCyberThreatIntelligenceBased2022,christinDissectingOneClick2010,lepochatPracticalApproachTaking2020b,maassEffectiveNotificationCampaigns2021}\\

    \end{tabular}
    \label{tab:selected_fields}
\end{table*}

\subsubsection{Parsing WHOIS.}

Parsing WHOIS data and extracting all pertinent fields presents a challenge, as detailed in Section~\ref{sec:background}. Consequently, this study focuses on specific fields used in previous research (see Table \ref{tab:selected_fields}), using custom templates designed to accurately parse various formats.
We developed 242 custom templates comprising regular expressions that outline the extraction process for selected fields from WHOIS records across numerous registrars. The templates are designed to handle multiple languages and formats, maximizing the comparability of records.

\subsubsection{Parsing RDAP.}
\label{paragraph:parsing-rdap}
Contrasted with WHOIS, parsing RDAP records is typically more straightforward, primarily due to the JSON format. Nevertheless, despite the data format being defined in \rfc{9083}, there %is often 
might be ambiguity regarding the correct placement of information within the data structure. Consequently, different registries and registrars may have varying interpretations of where specific information should be located.

We gathered the designated fields from all locations allowed by the RFC. %We collected the selected fields from all locations permitted by the RFC.
We considered malformed fields, those containing incorrect data types, or located in the wrong place within the data structure as missing. 
For instance, there are two primary representations of domain names in RDAP: as a string object (e.g., \texttt{ns.example.com}) or as an array of labels (e.g., \texttt{[ns, example, com]}). However, according to \rfc{9083}, when listing domain nameservers, they must be in the string format. Therefore, if we encountered a nameserver in the array format instead of the expected string format, we considered it as missing. This decision was based on the assumption that most automated systems would adhere to the RFC and disregard the field due to its invalid type. 
%, rendering the data unreliable for collection using automated tools.

\subsubsection{Field selection. \label{sussubsec:fields}}
To compare different data sources, it is important to note that not all registration data records share the same set of fields. As a result, we selected a limited number of fields, which have been commonly used in previous security studies and are consistently present in both WHOIS and RDAP records, whether at the registry or registrar levels. Table~\ref{tab:selected_fields} presents the selected fields, along with the type of data they hold and the articles that have used them. For this research, we have chosen the following fields:

\begin{itemize}

    \item \texttt{Nameservers}: this field indicates the name servers that have the authority over a particular domain.
    \item \texttt{IANA ID} and \texttt{Registrar}: the sponsoring registrar responsible for managing the domain is captured in the \texttt{Registrar} text field. Additionally, the \texttt{IANA ID} is an integer field that typically represents the unique identifier assigned by IANA~\cite{ianaRegistrars} to each ICANN-accredited registrar (if applicable).
    \item \texttt{Creation date} and \texttt{Expiration date}: these fields denote the date of the initial registration for the domain and the subsequent expiration date. Once the registration expires, the domain becomes available for purchase again unless the owner renews it.
    \item \texttt{Emails:} This field contains a range of contact email addresses that can be used, for instance, for reporting domain-related abuse.
\end{itemize}

We deliberately omitted selecting fields associated with a registrant, despite their use in several studies, due to their absence in many registries. Furthermore, the implementation of the European General Data Protection Regulation (GDPR) resulted in the removal or redaction of the field content by most servers. The impact of GDPR on the content of these fields falls outside the scope of this paper and has already been analyzed in prior research~\cite{luWHOISWHOWASLargeScale2021}.

% \todo{Remove this paragraph ?}
% Some fields were easier to parse than others. For example, the position of the \texttt{IANA ID} in the RDAP structure is
% well-defined, there are not many ways to spell the name of the field in WHOIS entries, even in
% different languages, and the content of the field should be an integer.
% On the other hand, parsing the different dates is challenging: for RDAP, \rfc{9083} imposes all date objects 
% to follow the well-defined format described in \rfc{3339}, but no such uniform format exists for WHOIS
% entries, so registrars use a very wide range of formats, sometimes using months written in the local registrar
% language or with custom formatting for time zones.
% For many registrars, a manual analysis was needed to infer the regular expressions that can parse a given date format.

When a field is absent from a record, or the content could not be parsed, the data is marked as missing. Table~\ref{tab:selected_fields} shows the proportion of records missing each field. 
%The record missing rate refers to the proportion of records for which the data is missing, while the domain missing rate represents the percentage of domains that have at least one record with missing data, considering that each domain has multiple records.
The record missing rate indicates the proportion of records with missing data, whereas the domain missing rate represents the percentage of domains that have at least one record with missing data, considering that each domain has multiple records (i.e., WHOIS and RDAP, including records collected by following referrals).

The missing rates for all fields, except for the IANA ID and Emails fields, are relatively low. This result was expected since the IANA ID solely pertains to domains under generic TLDs and ICANN-accredited registrars. 
%, and a significant number of email addresses in registration data were redacted or removed following the implementation of GDPR~\cite{luWHOISWHOWASLargeScale2021,icannTemporaryGTLD}.
%
Furthermore, each field presented its own set of parsing challenges, particularly in the case of WHOIS records, but also for RDAP. In RDAP, certain records, such as email contact addresses, can be located in different parts of the JSON structure as defined by \rfc{7483}.

\subsection{Analyzing Data Consistency}\label{subsec:mismatching_analysis}

After collecting, parsing, and cleaning the registration data for all studied domains, we analyzed the consistency among various WHOIS and RDAP records.
%In each collected entry, a field can either be present or missing. As described previously, malformed fields are considered to be missing as they can not be reliably detected and parsed.

For a given domain, if we were able to collect registration data from multiple sources and if these records have
common fields, we evaluated the consistency of the data. 
If the formatted data in same fields is identical, we considered them to be matching fields. 
On the other hand, if there is a discrepancy between the data, it results in a mismatch. 
We consider two types of mismatches: the first one involves two records from the same protocol, such as the registry WHOIS not aligning with the registrar WHOIS. The second type involves two records from different protocols, for instance, the registrar WHOIS not corresponding to the registrar RDAP.

\subsection{Ethical Consideration}

We adhered to the best practices recommended by the measurement community to ensure reliable results with minimal disruption to the servers~\cite{considerations,menlo}.
When gathering various data sources, including WHOIS, RDAP, and DNS records, we meticulously adhered to server rate limits~\cite{izhikevichZDNSFastDNS2022}. Additionally, upon visiting the scanner's source IP address, users are presented with a webpage that provides information about our identity, work, and instructions for adding a scanned server to our opt-out lists, allowing them to cease receiving requests from us. Throughout the study, we did not receive any opt-out requests via email.

The raw data we collected may include information about registrants. However, after the implementation of GDPR, most registrars provide options for their customers to choose which fields are visible or automatically redact personal information. In practice, most fields that could potentially contain personal data were redacted by default. 

\section{Results}

In this section, we present the %in-depth 
analysis of inconsistencies and explore the root causes of the disparities observed in specific fields.
Table~\ref{tab:selected_fields} provides a %thorough 
breakdown, field by field, indicating the count of records where the field was missing, the number of domains in which at least one mismatch was identified, or if the field was entirely absent from the records. Excluding the \texttt{emails} field, which raises its unique challenges discussed in Section~\ref{sec:emails}, we observed that 7.6\% of all examined domains exhibited at least one inconsistency in the remaining fields.

\subsection{Nameservers}\label{subsec:nameservers}

%The nameserver of a domain functions as the authoritative server responsible for providing DNS resource records for the domain, covering its zone and potentially subdomains if they are not delegated to other nameservers.
%
%When users attempt to resolve a specific DNS resource record for a domain, such as retrieving the IPv4 address of the domain via its \texttt{A} record, they initially need to determine the authoritative nameserver for the corresponding entries.

%Typically, the preferred method involves recursively querying the DNS tree, starting from the root zone and progressing towards the registry nameserver, which then provides the information about the nameservers managing the queried domain.
The typical method to obtain a list of authoritative nameservers for a given domain involves sending recursive queries within the DNS tree, starting from the root zone and progressing toward the registry nameserver, which then provides the relevant information~\cite{duEverchangingLabyrinthLargescale2016}.
However, in certain prior studies that had a primary focus on detecting malicious domains~\cite{christinDissectingOneClick2010,felegyhaziPotentialProactiveDomain2010,haoPREDATORProactiveRecognition2016}, the nameserver information used in the analysis was obtained from WHOIS.

The primary purpose of the nameserver fields was either to cluster domains with identical nameservers~\cite{christinDissectingOneClick2010,felegyhaziPotentialProactiveDomain2010} or to conduct further analysis on the nameserver itself. For instance,  investigations could involve verifying whether the nameserver is self-hosted, such as \texttt{ns.example.com} being authoritative for \texttt{example.com}, determining if it is %hosted by well-known generic nameservers,
managed by well-kown DNS service operators, or identifying if the apex domain of the nameserver is newly registered~\cite{haoPREDATORProactiveRecognition2016}.

In the subsequent part of this section, we begin by examining the various types of nameserver mismatches and their frequency. Then, we use DNS as a reference point %ground truth 
to ascertain the accuracy of the data sources involved in cases of mismatches.

%To analyze the nameservers mismatches, tags present at the end of lines are removed (like \texttt{ns.example.com [OK]}), then all nameservers are converted 
%to lowercase and trailing dots are removed, following \rfc{5890} stating that domain names can
%use a mix of upper-case and lower-case characters interchangeably and \rfc{9083} making trailing dots optional.

\begin{table}
    \center
    \caption{Number of records and domains with mismatching nameservers}
        \vspace{+0.4cm}
    \begin{tabular}{r|c|c}
        Case & Records & Domains \\
        \hline
        All & 1,044,268 & 576,204 \\
        Inclusion & 314,633 (30.1\%) & 224,833 (39.1\%) \\
        Intersection & 48,693 (4.6\%) & 23,934 (4.1\%) \\
        Disjoint & 680,942 (65.2\%) & 343,994 (60.0\%) \\ 
    \end{tabular}
    \label{tab:missmatch_ns}
\end{table}

\subsubsection{Mismatch Types.}

%One of the most common format errors when parsing nameservers from RDAP was encoding the nameservers as a list of labels, where \rfc{9083} clearly specifies that the \texttt{ldhNames} keys, containing the nameservers, must be in the string format.
%After cleaning the data, we detected 1,044,268 mismatches between two entries of the same domain, spread 
%over 576,204 unique domain names, which represents 1\% of the total collected domains.
We identified a total of 1,044,268 mismatches between two registration records of the same domain,  %but collected at different locations, 
encompassing 576,204 unique domain names. This accounts for approximately 1\% of the overall collected domains; hence 99\% of the measured domains did not have mismatching nameservers records.

When the nameservers of two records (referred to as $A$ and $B$) are found to be inconsistent, three potential scenarios may arise:

\begin{description} 
    \item[\textit{Inclusion.}] $A \subset B$ or $A \supset B$: one set is a subset of the other one. 
    \item[\textit{Intersection.}] No inclusion but $A \cap B \neq \emptyset$: $A$ and $B$ do not match but they have at least one common server.
    \item[\textit{Disjoint.}] $A \cap B = \emptyset$: $A$ and $B$ do not have common nameservers.
\end{description}

Table~\ref{tab:missmatch_ns} presents the number of mismatches detected in each scenario.
As described in Section~\ref{subsec:mismatching_analysis}, a given domain may have multiple records for each protocol, as each registration record may contain a referral field.
%As a consequence, each domain can have multiple types of mismatches: for instance, if the registrar WHOIS record is included in the registrar RDAP record, while the registry WHOIS record may also be disjoint from the registry RDAP record, this domain would count both in inclusion and disjoint.
%Therefore, the values on the domain column add up to more than 100\%.
As a result, each domain can exhibit multiple types of mismatches. For example, the nameservers extracted from the registrar's WHOIS record could be included in the list of nameservers found in the registrar's RDAP record, and additionally, the nameservers listed in the registry's WHOIS record could entirely differ from the servers in the registry's RDAP record. In such cases, a domain would be counted in both the inclusion and disjoint categories. Consequently, the values in the Domains column may exceed 100\%.

%When using DNS to find the resource records of a domain, if the client (e.g., recursive resolver) has a choice between multiple nameservers, it can use any
%of them interchangeably, or even query all of them and process the first received answer~\cite{rfc1035}.
%As a consequence, the inclusion and intersection cases %should 
%might be less concerning because both records share at least one 
%nameserver, which could be a sign that all nameservers provided by both records serve the same data.
%On the contrary, the disjoint case for which both records have no server in common is  concerning as it raises the 
%suspicion that both nameservers may not serve the same data or be authoritative for a given domain name.
When using DNS to fetch the domain's resource records, if the client (e.g., a recursive resolver) has multiple nameservers to choose from, it can use any of them interchangeably or query all of them and process the first received answer~\cite{rfc1035}. This means that the inclusion and intersection cases may be less concerning, as both records share at least one nameserver, potentially indicating that all nameservers serve the same data. Conversely, the disjoint case, in which both records do not have common servers, is concerning as it raises suspicion that the nameservers may not serve the same data or be authoritative for the domain name.
This %problematic
situation concerns 65\% of the studied mismatches and 60\% of the domains with mismatching records. 
The mismatch often involves records from different protocols. We have observed that 67.6\% of the nameserver mismatches were between a 
WHOIS record and an RDAP record, whereas 17\% were between two RDAP records (registry RDAP and registrar RDAP) and 
15.4\% were between two WHOIS records of the same domain.

%To conclude on this initial analysis of nameserver mismatches, even if it only concerns 1\% of the domains, the types 
%of mismatches raise concerns: 67.6\% of mismatches involve a difference between WHOIS and RDAP and in 60\% of the cases, both
%entries have no nameservers in common. As a consequence, when gathering nameserver information on a domain, the choice 
%between WHOIS and RDAP is not neutral and will yield different incompatible results.

In summary, while affecting only 1\% of domains, nameserver mismatches, especially the 67.6\% involving disparities between WHOIS and RDAP, raise concerns. In 60\% of such cases, both sources lack any common nameservers, making the choice between WHOIS and RDAP for gathering nameserver information non-neutral and yielding incompatible results.

\subsubsection{Who is Right?}\label{subsec:who_is_right}

To successfully collect any DNS record for a domain %to be successfully resolved to an IP address, 
it is essential to have an \texttt{NS} record in the parent zone file, specifying the authoritative nameserver for the domain. To gather the nameserver information, we actively queried the DNS infrastructure and performed a comparison with the nameservers listed in the WHOIS and RDAP records.

\paragraph{\textbf{Methodology.}}\label{subsubsec:ns_methodo}

To find the \texttt{example.com} nameservers, the client (e.g., a recursive resolver) 
first sends 
an \texttt{NS} query to the DNS root servers and receives the name of the servers that have authority over the \texttt{.com} zone.
The client then sends another \texttt{NS} query to one of these servers and receives the \texttt{NS} record of \texttt{example.com}.
This last answer comes from the registry in charge of the \texttt{.com} zone.
%As pointed out in previous work~\cite{sommeseWhenParentsChildren2020}, t
The client %(recursive resolver) 
can then either return the result because it retrieved the \texttt{NS} record of \texttt{example.com} from %an authoritative server), 
the authoritative nameservers of the parent (nameserver of \texttt{.com})
or perform additional \texttt{NS} queries to the nameservers received at the previous step and get the 
nameservers configured by the administrator of the domain.
\rfc{1034} states that the nameservers returned by the registry and the nameservers configured by the administrator must be 
identical, but previous study~\cite{sommeseWhenParentsChildren2020} revealed that around 10\% of the domains in 
the \texttt{.com}, \texttt{.org} and \texttt{.net} zones had differences between the nameservers provided by the parent registry servers 
and the nameservers provided by the child domain servers.
If a domain is active, it must have an \texttt{NS} record at the registry level, as it is a part of the resolution chain.
On the contrary, some domain owners do not put \texttt{NS} records in the child nameservers.
To maximize the number of collected domains, we queried the \texttt{NS} resource records for each domain at the registry level.

\paragraph{\textbf{Scans.}}
%To find which registration data source is consistent with DNS data, we used  \texttt{zdns}~\cite{izhikevichZDNSFastDNS2022} to collect the \texttt{NS} resource record of each domain for which 
%we have detected a mismatch, and collected their WHOIS and RDAP entries a second time, between January 24th and January 27th 2023
%to make sure that all three data sources 
%(DNS, and the two mismatching entities) were collected at the same moment, and avoid cases for which the domain configuration 
%was modified during our scans.
%Some domains expired between our first scan and this additional analysis, so we were not able to check all the detected mismatches
%but 90\% of domains were still active and returned \texttt{NOERROR} and a non-empty result when scanned.
To determine the consistency between registration data sources and DNS data, we used \texttt{zdns}~\cite{izhikevichZDNSFastDNS2022} to retrieve the \texttt{NS} resource records of each domain where a mismatch was detected. Additionally, we collected their WHOIS and RDAP records for a second time, specifically between January 24th and January 27th, 2023, which ensured that all three data sources (DNS, WHOIS, and RDAP) were collected simultaneously, eliminating cases in which domain configurations were altered during our scans.

While some domains had expired between our initial scan and this supplementary analysis, approximately 90\% of the domains remained active and returned a \texttt{NOERROR} DNS response with non-empty results during the scan.

\paragraph{\textbf{Results.}}

%The first analysis of nameservers detected 576,204 domains with mismatching entries.
%The new data collection and analysis revealed that 37\% of the scanned domains did not present mismatches anymore, but still detected 365,521 unique domains with mismatching nameservers.

The second data collection unveiled 365,521 distinct domains exhibiting nameserver mismatches.

After the collection of the new registration data and the \texttt{NS} records from the authoritative DNS servers, the resulting
data falls into two categories: the mismatch can be between two records from the same protocol (two WHOIS records or
two RDAP records), or between two records of different protocols.

\begin{figure}
    \includegraphics[width=1\textwidth]{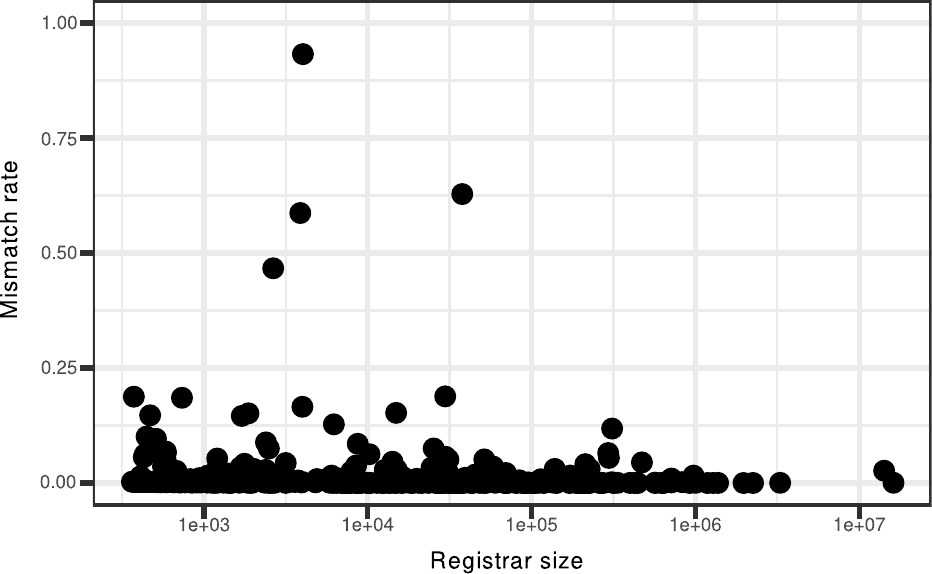}
    \centering
    \caption{Nameserver mismatch rate per registrar}
    %\vspace{-0.3cm}
    \label{fig:ns_iana}
\end{figure}

\paragraph{WHOIS-RDAP mismatches.}
In 74.9\% of the identified mismatch cases, the disparity exists between a record gathered through WHOIS and a record collected through RDAP. As previously described, the nameservers obtained from DNS may constitute a subset, superset, or have a non-empty intersection with each record.
Upon examining all possible scenarios, we found that in 99.5\% of cases, the DNS record corresponded to either the WHOIS or RDAP record. The remaining 0.5\% involved intermediate situations in which the DNS result only partially matched one of the records. Due to the limited number of domains affected by this situation, we opted for concentrating our analysis on cases in which the DNS matched one of the records.

In 78.5\% of cases, the DNS data corresponded to the nameservers provided by the RDAP record. This underscores the fact that, although nameservers obtained from DNS typically align with data from RDAP, there are still 21\% of mismatch instances in which the DNS results match the WHOIS record.
Interestingly, Figure~\ref{fig:ns_iana} highlights that a few registrars exhibit a notably high mismatch rate compared to others. We observed that only four registrars have a mismatch rate exceeding 25\%, while the largest registrars, representing the majority of domains, maintain a very low mismatch rate.

\paragraph{Registry-Registrar mismatches.}
The remaining 25.1\% of cases represent the situations in which the mismatch is between two records from the same protocol but collected
from different servers. 
In this case, the collector queried the registry server, got a referral to another server, and recursively called it, gathering
an additional record.
If two records are inconsistent, we checked if the nameservers provided by the DNS matched the records collected at the registry server
or at the referral servers. 
In 99.2\% of the cases, the DNS data matched the registry record, and in the remaining 0.8\% of the cases, it did not match either
records. 
The DNS data matched the registrar record in only 0.008\% of the cases.

As described in Section~\ref{subsubsec:ns_methodo}, we decided to collect the \texttt{NS} records at the DNS authoritative nameservers 
of the registry.
Consequently, we expected the record provided by the registry to be consistent with the DNS data from the same registry.
Hence, the mismatches between two records from the same protocol almost always come from invalid data from the referral server.

%The main takeaways of this analysis on nameservers are that 1\% of domains present two records with mismatching nameservers.
%In the majority of these cases, both sets of nameservers have nothing in common and when the difference is between
%an RDAP and a WHOIS record, the RDAP record is right and matches the \texttt{NS} records from DNS in 78\% of the cases.
The main takeaway is that when both sets of nameservers do not have common elements, and the discrepancy lies between an RDAP and a WHOIS record, the RDAP record is accurate and aligns with the \texttt{NS} records from DNS in 78\% of the cases.

\subsection{IANA ID, Creation and Expiration Dates}\label{subsec:creation_expiration_dates}

When it comes to obtaining the IANA ID, creation date or registrar name of a domain, research primarily relies on the WHOIS and RDAP protocols. Unlike nameservers, which can also be retrieved from DNS, there is no third-party service that offers direct access to this data. Consequently, when two sources diverge in these fields, there is no simple method to determine which record contains the accurate information.

In this section, we outline the types of mismatches identified in IANA ID, creation and expiration dates, and highlight a few cases in which we can ascertain the correct record.

\subsubsection{Creation and Expiration Dates.}
The creation date represents the domain's initial registration instant, providing insight into its age. In domain-related research, the domain age is a pivotal factor as older domains, active for multiple years, are generally deemed more trustworthy than newly registered ones. The extensive analyses of the domain registration behavior~\cite{affinitoDomainNameLifetimes2022,haoPREDATORProactiveRecognition2016} have shown that malicious domains tend to have shorter lifespans and are used in attacks shortly after registration. 
Other studies~\cite{duEverchangingLabyrinthLargescale2016,felegyhaziPotentialProactiveDomain2010} have used the creation date to detect bulk registrations of malicious domains.

The domain age is also frequently combined with other parameters to distinguish between benign and malicious domains~\cite{ghalebCyberThreatIntelligenceBased2022,lepochatPracticalApproachTaking2020b}. While some approaches~\cite{affinitoDomainNameLifetimes2022} attempt to estimate the domain activity period by monitoring its appearance and disappearance in publicly accessible zone files, this method is contingent on zone file accessibility and the availability of historical data for the domain. Consequently, most studies depend on WHOIS or RDAP to acquire the creation date.

\begin{figure}
    \center
    \includegraphics[width=0.8\textwidth]{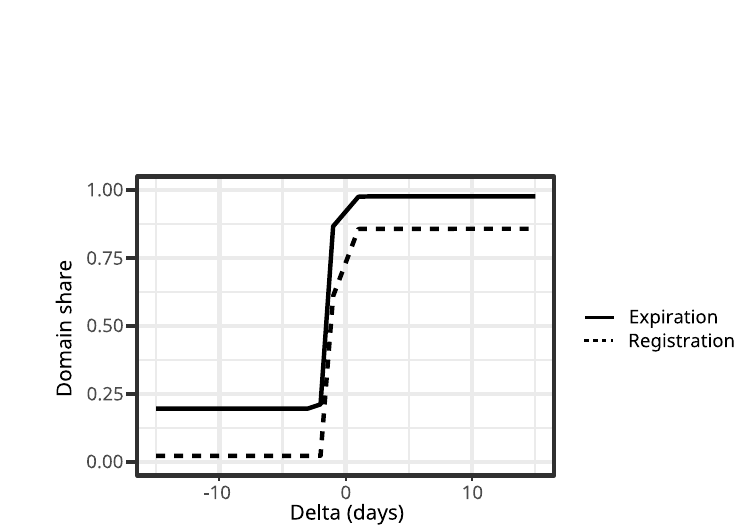}
    \caption{Cumulative distribution of creation and expiration date mismatches}
    \label{fig:creation_expiration_date}
\end{figure}

The expiration date also provides insights into the domain behavior and can shed light on various scenarios. For instance, if a domain is removed from its zone file before its expiration date, it may suggest actions taken by the registrar or seizure by authorities~\cite{affinitoDomainNameLifetimes2022}. Additionally, parking and drop-catching entities use the expiration date to identify when a domain will become available 
for re-registration~\cite{vissersParkingSensorsAnalyzing2015}.

Both creation and expiration dates are usually found in the majority of WHOIS and RDAP records. However, in the case of WHOIS, they may be listed under various names, such as \texttt{Creation Date}, \texttt{Registration Date}, \texttt{Created at}, \texttt{Valid until}, and more.

After filtering out the dates that were not possible to parse and the dates lower or equal to the UNIX Epoch (which may indicate a default value or a configuration error), we observed that 5.7\% (for creation dates) and 4.4\% (for expiration dates) of the domains exhibited inconsistencies across their records. Figure~\ref{fig:creation_expiration_date} illustrates the distribution of time differences between these records.

We can observe that in 84\% of the cases for creation dates and 78\% of the cases for expiration dates, the differences are less than 2 days. These discrepancies have minimal impact on the analyses relying on creation dates to gauge the domain age~\cite{haoPREDATORProactiveRecognition2016} or on the speed of domain re-registration after expiration~\cite{affinitoDomainNameLifetimes2022}.

\begin{figure}
     \includegraphics[width=1\textwidth]{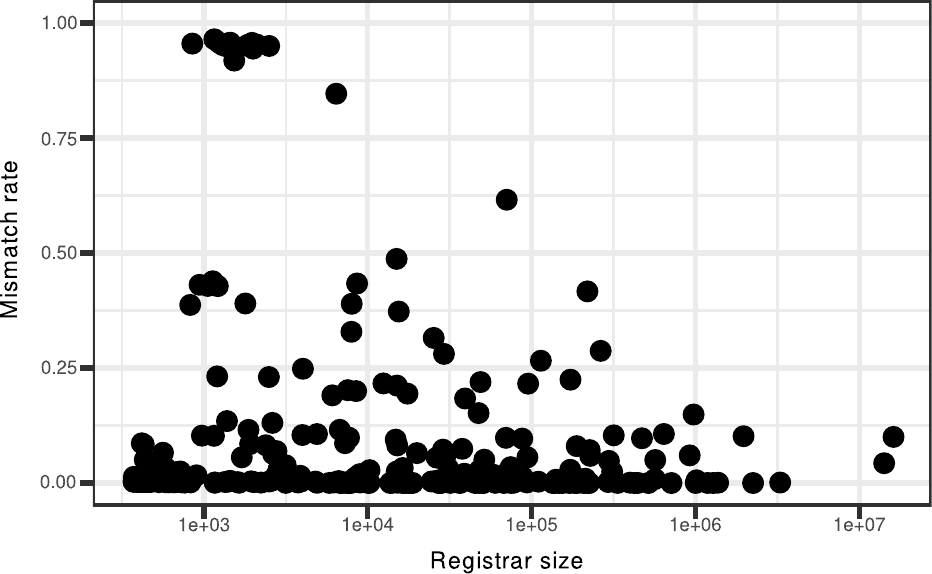}

    \centering
    \caption{Creation date mismatch rate per registrar}
    \label{fig:creation_mismatch}
\end{figure}

Previous studies~\cite{lauingerWHOISLostTranslation2016} highlighted common misunderstanding
of the different expiration steps before the deletion of a domain and pointed out that these steps can 
account for a mismatch of up to 30 days, as a confusion could be made between the expiration date, the deletion date 
and how the grace and redemption periods should be accounted for, but the collected data shows no specific mismatch proportion 
at 30 days. 
However, our analysis points out that several records present an expiration date difference of exactly one year, which corresponds
to the minimal duration of a registration, so the difference could come from the fact that the renewal of the domain was taken into 
account in one of the records and not in the other.
Then, 98\% of expiration date mismatches are either under 2 days or exactly 1 year, leaving only a few domains with unexplained 
expiration date mismatches.

\begin{comment}
On the other hand, more than 16\% of the creation date mismatches are over 2 days, and, contrary to the expiration date 
mismatches, the creation date differences over 2 days are evenly spread, providing no technical explanation on the source of such mismatches.
This absence of clear behavior raises the suspicion that entities may not have the same definition of the \texttt{Creation Date}
because even if \rfc{9083} clearly defines keywords to describe creation events in RDAP records and makes the distinction
between \texttt{registration, reregistration, reinstantiation}, and \texttt{transfer}, it is not the case for WHOIS, 
so the \texttt{Creation Date} filled in the WHOIS record and the \texttt{registration} event in the RDAP record may not represent the 
same events in the domain life-cycle.
The \texttt{Creation Date} mismatch rate of each registrar, represented in Figure~\ref{fig:creation_mismatch}, highlights that 
even if many registrars have over 10\% of their domains with mismatches, a few registrars have almost 100\% of their
domains with the mismatching creation date, confirming our assumption that mismatches may come from registrar configuration mistakes.
\end{comment}

Approximately 16\% of the creation date mismatches extend beyond 2 days. In contrast to expiration date mismatches, creation date mismatches are more evenly distributed.
One possible explanation for these discrepancies is that different entities may have distinct definitions of the \texttt{Creation Date}. While \rfc{9083} clearly defines keywords to describe creation events in RDAP, such as \texttt{registration},  \texttt{reregistration}, \texttt{reinstantiation}, and \texttt{transfer}, WHOIS lacks such precision. Consequently, the \texttt{Creation Date} recorded in the WHOIS record may not correspond to the same events in the domain life cycle as the \texttt{registration} event in the RDAP record.

The \texttt{Creation Date} mismatch rate for each registrar, as shown in Figure~\ref{fig:creation_mismatch}, highlights that while many registrars have over 10\% of their domains with creation date mismatches, a few registrars exhibit nearly 100\% of their domains with mismatched creation dates. This observation supports our hypothesis that some of these mismatches may result from registrar misinterpretations, custom registration processes, or systematic configuration errors.
For example, the vast majority of domains presenting a \texttt{Creation Date} mismatch of 30 or 31 days are under the \texttt{.com} TLD and share the same registrar, \texttt{FastDomain Inc}. For these domains, the registrar record \texttt{Creation Date} is always one month earlier that the one in the registry record. After investigation, we found that this registrar allows their customers to cancel their domain order up to 30 days after payment, while the ICANN Agreement~\cite{icannAgreement} only imposes a 5-day refund window. Consequently, we can hypothesize that the creation of the registry record was delayed until the end of the 30-days period, while the registrar record was created when the customer first ordered the domain.

\begin{comment}
The Creation and Expiration Dates are crucial for many security metrics as they are important information on the domain
life cycle.
However, 5.4\% of domains provide multiple inconsistent results with no easy way to check the extracted data against another data source.
Even if the number of affected domains is relatively low, some registrars have a very high mismatch rate. This raises suspicion on the
quality of the data they provide, the confidence that the community can place in the dates extracted from their records, and therefore, on
the effectiveness of systems relying on the precision of this data.
\end{comment}

\begin{table}
    \center
    \caption{Number of records and domains with mismatching emails}
    \vspace{+0.4cm}
    \begin{tabular}{r|c|c}
        Case & Records & Domains \\
        \hline
        All & 50.1M & 19.0M \\
        Inclusion & 37.1M (74\%)& 15.1M (79.8\%) \\
        Intersection & 0.59M (1.2\%) & 0.56M (2.9\%) \\
        Disjoint & 12.4M (24.8\%) & 4.9M (26\%)\\
    \end{tabular}
    \label{tab:mismatch_mails}
\end{table}

\subsubsection{IANA ID and Registrars.}\label{subsec:iana_id_registrars}
\begin{comment}
Most registries in charge of Top Level Domains (TLDs) do not directly sell the domains they manage, instead they
delegate this task to registrars.
Most of them are accredited by ICANN, the organization that manages the assignment of TLDs to registries and allows registrars
to sell domains under Generic TLDs (gTLDs) on the condition that they follow the Registrar Accreditation 
Agreement~\cite{icannAgreement}. 
Once accredited, the registrar receives a unique registrar identifier, the \texttt{IANA ID}, and ICANN updates
their list of accredited registrars~\cite{ianaRegistrars}, mapping the registrar name to its ID.
This \texttt{ID} is used to uniquely identify registrars and is present in the vast majority of WHOIS and RDAP records, 
even for domains under a Country Code TLD (ccTLD) that do not require registrars to be accredited to sell their domains.

As a consequence, all gTLDs and most ccTLDs WHOIS and RDAP records have a \texttt{Registrar} field containing the name of the registrar that manages
the domain, and an \texttt{IANA ID} field, containing the unique ID of this registrar.
These values can change over time because domains can be transferred from one registrar to another. 
When it happens, the registry must update the referral server of the domain so that it points to the new registrar server,
and update the \texttt{Registrar} and \texttt{IANA ID} fields.
\end{comment}

ICANN-accredited registrars play a crucial role in domain registration and management. The IANA ID associated with each registrar is a unique identifier, often found in WHOIS and RDAP records, helping to trace domain ownership and authority.

\begin{comment}
It is important to note that the content of the\break \texttt{Registrar} field often does not match the name registered in the
IANA list.
Registrars tend to use many different ways of writing their names, and the notation changes over time.
For example, according to the IANA, the registrar with ID number 146 is \texttt{GoDaddy.com, LLC}, but we found 
that 2.4\% of the domains with this IANA ID (385,033 unique domains) have a different \texttt{Registrar} field, like
\texttt{GoDaddy}, \texttt{GoDaddy LLC} (without the \texttt{.com}) or \texttt{GoDaddy LLC.} (with a trailing dot), and we detected 7 different
orthographies for this registrar name only.
As a consequence, parsing the \texttt{Registrar} field to detect the exact registrar is a difficult process and users prefer 
parsing the \texttt{IANA ID}~\cite{haoPREDATORProactiveRecognition2016,felegyhaziPotentialProactiveDomain2010}. 
However, some ccTLDs do not use accredited registrars and IANA IDs, so in these cases, extracting registrar information can
only be done through the \texttt{Registrar} field.
\end{comment}

The content of the \texttt{Registrar} field in WHOIS and RDAP may differ from the name listed in the IANA registry. For example, 2.4\% of domains with IANA ID 146 (\texttt{GoDaddy.com, LLC}) have different \texttt{Registrar} entries, including  \texttt{GoDaddy LLC},  \texttt{GoDaddy.com, Inc.}, \texttt{GODADDY} or \texttt{Go Daddy, LLC}. Therefore, parsing the \texttt{Registrar} field to identify registrars can be challenging, and users often rely on the \texttt{IANA ID} for accuracy. However, in certain ccTLDs, registrars receive local accreditation, and the corresponding IANA IDs are not assigned or displayed in the public WHOIS and RDAP. In these cases, extracting registrar information solely relies on the \texttt{Registrar} field.

%We only expected ID mismatches in a few cases in which the domain is transferred between different registrars and the 
%modification of the ID was not done in both WHOIS and RDAP yet.
%Our analysis detected that 0.2\% of domains had records with mismatching \texttt{IANA ID}
%, which corresponds to what was 
%expected as the field is easy to parse and well-defined, both for WHOIS and in \rfc{9083} for RDAP, and
%registrar transfers are not very common.
Our analysis uncovered that a mere 0.2\% of domain names had records with inconsistent \texttt{IANA ID}.
The analysis of IANA IDs reveals that the majority of mismatches occur between specific pairs of IDs. Approximately 91\% of these detected mismatches involve a record with IANA ID 1556 (\texttt{Chengdu West Dimension Digital Technology Co., Ltd.}) and another record with IANA ID 1915 (\texttt{West263 International Limited}). Additionally, 4\% of the mismatches involve IANA ID 3951 (\texttt{Webempresa Europa, S.L.}) and ID 5555555, which is an invalid ID. This pattern may suggest misconfiguration issues by particular entities, resulting in consistent mismatches across all the domains they manage.

In the second case, we confirmed the issue by registering a domain name  %gegxkkuyaz.com 
with the registrar \texttt{Webempresa Europa, S.L.} and examining its records. While the registry WHOIS record correctly indicated the valid IANA ID 3951, the registrar WHOIS record contained an IANA ID field with the value 5555555, which does not correspond to any valid registrar number. 
The registrar's WHOIS record also displayed placeholder values for various fields, including the abuse contact phone number and the reseller name. We verified that all domains registered with this registrar had inconsistent records. We reported the issue to the registrar, and over several months, we noticed that all the domains they managed were updated with correct registration data, resolving the inconsistencies.
%The registrar did not provide any additional explanation about this initial misconfiguration.
We suspect that the mismatches between ID 1556 and ID 1915 share the same origin. However, we were unable to test this hypothesis, as both registrars exclusively serve users in China and Hong Kong.

%IANA ID is a crucial part of registration data as many domain detection systems use it to group the domains managed by the 
%same registrar, as some behaviors may be shared. The \texttt{Registrar} field from WHOIS and RDAP is impractical to use as each
%registrar can write its name in different ways. Therefore, security researchers prefer using the IANA ID field that
%applies to the vast majority of domains and is easy to parse. This field has the lowest mismatch rate detected in this 
%study, concerning only 0.2\% of the domains. However, we still observed errors, likely due to registrar misconfigurating
%their records, or using the field for internal purposes.

\subsection{Email Addresses}\label{sec:emails}

\begin{comment}
Several different types of email addresses can be put in the registration data: contact addresses linked to the registry, 
registrar, or registrant, to contact these actors for generic purposes, and tech, admin, and abuse addresses used to declare 
configuration problems, report illegal abuse or fill complaints.
\rfc{9083} defines many different keywords in RDAP that can be used to describe the precise role of a mail address, like 
\texttt{administrative}, \texttt{abuse}, \texttt{billing}, or \texttt{technical}, so it is relatively easy to find the role of the addresses.
On the contrary, in WHOIS records, because of the lack of the structure, email addresses can be put anywhere, and a precise comprehension
of the context is needed to determine their exact role. 
%Moreover, many registrars do not detail the role of each email address in their WHOIS entries, probably expecting users to infer 
%the role of the address from its syntax, for example, considering that \texttt{abuse@registrar.com} is 
%the abuse email address.
%However, no specification describes this notation and some registrars have their own naming conventions, leading to the situations in which
%the role of the email is impossible to determine, for example, it is impossible to know if \texttt{contact@registrar.com} should be 
%used for an abuse report or if \texttt{support@registrar.com} should be preferred.
\end{comment}

Various types of email addresses are included in registration data, serving different purposes. These addresses are associated with the registry, registrar, or registrant, as well as for technical, administrative, and abuse-related functions. \rfc{9083} provides specific keywords in RDAP for describing the role of each email address, such as \texttt{administrative}, \texttt{abuse}, \texttt{billing}, or \texttt{technical}, which allows for easy identification of the address role, a capability that WHOIS lacks.
%
\begin{comment}
Detecting the role of each email address present in a record requires a lot of empirical and manual categorization, so we gathered all addresses present in
each record without any role distinction, and compared the different records based on the sets of addresses they contain.
We expected mismatches due to the possible presence of protocol-specific contact addresses, for example, if the registrar
delegates the technical administration of their RDAP servers to a third-party company, the technical contact email
could be different from the technical contact mail for the WHOIS records.
However, we expect some addresses to be in common between the multiple records of a same domain, like the abuse contact email for 
contacting the managers of a domain to report abuse linked to the domain. 
%This address is not protocol-dependant, hence, it should be the same across all entries, servers and protocols.
\end{comment}
%
For these reasons, we chose to collect all addresses in each record without distinguishing their roles. We then compared the records based on the sets of addresses they contain. Mismatches can occur due to protocol-specific contact addresses; for instance, the technical contact email for RDAP records may differ from that in WHOIS records if a registrar delegates technical administration to a third party. However, we anticipate that some addresses will be common across multiple records for the same domain, such as the abuse contact email for reporting domain-related abuse.

%To analyze the email mismatches, we used the same techniques as described for nameservers in Section~\ref{subsec:nameservers}.
%First, email addresses are parsed and deduplicated, and the different possible inclusion and intersection cases are compared.
%Table~\ref{tab:mismatch_mails} presents the results of this analysis.

To analyze email mismatches, we applied the techniques described in Section~\ref{subsec:nameservers}. Initially, email addresses were parsed and duplicates removed. Subsequently, we compared the various possible inclusion and intersection cases. Table~\ref{tab:mismatch_mails} presents the results of this analysis.

\begin{comment}
We detected 50 million mismatches on 19 million unique domains, covering 34.5\% of the domains studied in this work.
For most mismatches (74\%) and most domains (79.8\%), one set of email addresses was included in the other. 
The majority of mismatches (75.2\%) are Inclusions or Intersections and may correspond to the situations in which some addresses 
are shared (like the abuse or registrant email) while server or protocol-specific addresses are added by the different entities 
(like the contact address for the entity managing this specific server) and may differ.
However, almost 5 million domains (8.8\% of all the analyzed domains) had a pair of records that had absolutely no
email addresses in common.
\end{comment}

We identified 50 million mismatches for 19 million unique domains, encompassing 34.5\% of the domains in this study. Among them, 74\% of mismatches and 79.8\% of domains featured one set of email addresses included in the other. About 75.2\% of mismatches were either inclusions or intersections, potentially arising from shared addresses (e.g., abuse or registrant emails) while the addition of server or protocol-specific addresses by different entities (e.g., contact addresses for WHOIS or RDAP servers) may result in differences. However, nearly 5 million domains (8.8\% of all analyzed domains) had a pair of records with no common email addresses.

The disjoint cases may be attributed to the GDPR implementation. Previous research~\cite{luWHOISWHOWASLargeScale2021} explored the impact of GDPR on the availability of personal information fields before and after its enactment. 
Following the GDPR implementation, many registrars and registries replaced the registrants' personal details like the name, the phone number, and the email address in WHOIS and RDAP records with entries such as `REDACTED FOR PRIVACY', effectively concealing this information.  However, some entities introduced proxy email addresses to safeguard the registrants' actual addresses. These proxy servers mediate communication between proxy addresses and registrant emails. For example, in an RDAP record under the \texttt{registrant} role, one might encounter the address \texttt{b4ebaf9bfeba@withheld forprivacy.com}. While this conceals the registrants' personal data from the public, a valid contact address remains accessible.
%
%Protecting  user privacy by redacting their email address or by hiding them behind a proxy can introduce an 
%artificial difference between the different WHOIS and RDAP records because the registrant address that should have
%been in common in all records gets redacted or hidden behind a proxy that may not be the same for all records.
%
Protecting user privacy by redacting or using proxy email addresses can create discrepancies between WHOIS and RDAP records, as the registrant's address, which should be consistent in all records, may be redacted or hidden behind proxies.

Email mismatches can also occur when registrars or registries use distinct addresses for WHOIS and RDAP, even though both email addresses are administered by the same organization, such as \texttt{abuse.whois@registrar.com} and \texttt{abuse.rdap@registrar.com}.

\begin{comment}
To circumvent the artificial differences introduced by these two techniques, 
we re-run our data analysis after dropping the local part of the address, only keeping the base domain name.
In this way, email addresses hosted by the same domain are considered to be consistent.
Table~???\ref{tab:mismatch_mails_domains} shows the results of this new analysis.
We observe that this process resolved 18.6\% of the mismatches.
With this operation, the rate of Disjoint email addresses drops from 24.8\% to 9.7\%, effectively resolving 40\% of the
Disjoint situations. 
This argument validates the hypothesis presented before: most situations in which the email addresses were disjoint came from
records for which the email addresses were different but still hosted under the same domain.
\end{comment}

To address these discrepancies, we conducted a new analysis by extracting and comparing only the domain names from email addresses, discarding the local parts. This approach considered email addresses within the same domain as consistent. The results are presented in Table~\ref{tab:mismatch_mails_domains}. We found that this approach resolved 18.6\% of the mismatches and reduced the rate of disjoint email addresses from 24.8\% to 9.7\%, which suggests that in many cases in which email addresses appeared disjoint, they actually originated from records with different email addresses hosted under the same domain.

\begin{table}
    \center
    \caption{Number of records and domains with email domain mismatches after removing the local part of the address, retaining only the base domain name}
    \vspace{+0.5cm}
    \begin{tabular}{r|c|c}
        Case & Records & Domains\\
        \hline
        All & 50.1M & 19M \\
        Equality & 9.3M (18.6\%) & 4.0M (21.4\%) \\
        Inclusion & 35.7M (71.3\%)& 14.5M (76.7\%) \\
        Intersection & 0.24M (0.5\%) & 0.23M (1.2\%) \\
        Disjoint & 4.8M (9.7\%) & 2M (10.6\%)\\
    \end{tabular}
    \label{tab:mismatch_mails_domains}
\end{table}

\begin{comment}
This analysis of the email mismatches shows that collecting email addresses, for example for notification 
campaigns~\cite{maassEffectiveNotificationCampaigns2021}, should be done with extra care: for 34.5\% of domains, 
the collected emails will heavily depend on the data source used to gather the addresses.
Moreover, when entries mismatch on email addresses, the domains hosting the addresses are completely disjoint in 10\% of 
the cases.
For a given domain, these differences hint that the email servers are managed by completely different entities, and 
the effect of notification campaigns could be highly variable, depending on where the emails were collected, both 
in terms of the validity of the addresses as well as the responsiveness of the entity managing it.
\end{comment} 

%This analysis highlights the importance of exercising caution when collecting email addresses, particularly for notification campaigns~\cite{maassEffectiveNotificationCampaigns2021}. For 34.5\% of domains, the choice of the data source significantly impacts the collected email addresses. Furthermore, in 10\% of cases for which the email records mismatch, the domains hosting these addresses are completely unrelated, which indicates that the email servers are managed by distinct entities, introducing substantial variability in the effectiveness of notification campaigns. The validity of the addresses and the responsiveness of the managing entity depend on the sources of the emails.

In conclusion, this analysis underscores the need for caution when gathering email addresses, especially for notification campaigns~\cite{maassEffectiveNotificationCampaigns2021}. The choice of data source significantly affects the collected email addresses for 34.5\% of domains. Additionally, in 10\% of cases for which email records mismatch, the domains hosting these addresses are unrelated, suggesting that email servers may be managed by different entities, potentially leading to varying effectiveness in notification campaigns.

\section{Related Work}\label{sec:related-work}

Table~\ref{tab:selected_fields} provides an overview of prior research that used WHOIS and RDAP data for domain name registration information. Nevertheless, the accuracy of the collected data has not been thoroughly investigated. Some earlier studies~\cite{christinDissectingOneClick2010,duEverchangingLabyrinthLargescale2016,lauingerWHOISLostTranslation2016,felegyhaziPotentialProactiveDomain2010} relied on WHOIS data prior to the introduction of RDAP. However, as discussed in Section~\ref{sec:background}, inconsistencies are also present in WHOIS data obtained from servers managed by registries and registrars.

Challenges in processing WHOIS records have been identified, particularly concerning the reliability of extracted data such as AS numbers for IP WHOIS~\cite{bianzinoWhoWhoisAnalysis2014} and domain status~\cite{lauingerWHOISLostTranslation2016}. In a previous in-depth analysis of the \texttt{.com} zone~\cite{liuWhoComLearning2015}, the authors developed a machine-learning algorithm to address the multiple formats used in WHOIS records, demonstrating the difficulties in consistently parsing relevant fields.

The performance analysis of WHOIS and RDAP \cite{gananWHOISSunsetPrimer2021} focused on the speed but lacked the examination of data consistency across different servers and protocols.

In our work, we observed that 7.6\% of the scanned domains exhibited mismatching records, raising concerns about the reliability of security metrics relying on such data. Notably, metrics that use the \texttt{Creation Date} field \cite{lepochatPracticalApproachTaking2020b} or the bulk registration status \cite{affinitoDomainNameLifetimes2022} may be impacted, especially for registrars with high mismatch rates as presented in Figure~\ref{fig:creation_mismatch}. Obtaining accurate creation dates for domains under these registrars may require alternative data sources.

The \texttt{Emails} field exhibited the highest mismatch rates, even with a conservative parsing approach. Previous studies on notification campaigns \cite{maassEffectiveNotificationCampaigns2021,cetinUnderstandingRoleSender2016} reported difficulties in extracting valid email addresses from WHOIS records, with email bounce rates exceeding 50\%. These findings raise concerns about the effectiveness of notification campaigns due to the challenges associated with obtaining  
consistent and valid abuse emails from different entities.

\section{Conclusions}\label{sec:conclusion}

Registration data plays a crucial role in the development of detection systems and gaining insights into the domain name behavior and entity management. However, obtaining this information may require interacting with various servers (either registries or registrars) and protocols (either WHOIS or RDAP). Our extensive analysis of 164 million records from 55 million domains unveiled that the data obtained through WHOIS and RDAP is generally consistent.
Nonetheless, 7.6\% of the analyzed domains displayed discrepancies in one or more of the following fields: IANA ID, creation and expiration dates, or nameservers. 
%This underscores the variations observed across different servers and protocols. 
In cases related to the nameserver field, we used active DNS measurements to determine the accurate record. When disparities involved RDAP and WHOIS records, our findings showed that RDAP records were correct in 78\% of instances for which mismatches occurred.

%The key finding includes the need for studies relying on reliable registration data to collect it from different servers and protocols, corroborating values with third-party sources. While larger registrars generally exhibit lower mismatch rates, this evidence does not guarantee data validity. Smaller registrars show diverse results, with some having minimal mismatch rates while others exhibiting high rates. Malicious actors could exploit the registrars with inconsistent data to evade detection systems that rely on registration data availability and quality. Analyzing the number of malicious domains managed by inconsistent registrars would provide valuable insight into evasive tactics. To support future research, we will share the collected records (raw and parsed) and data analysis as artifacts associated with this publication.

The principal insight underscores the importance of studies that rely on dependable registration data to diversify their data sources by collecting it from various servers and protocols. 
Although larger registrars generally display lower mismatch rates, this observation does not inherently guarantee the accuracy of the data. Smaller registrars present a wide range of outcomes, with some demonstrating minimal discrepancies, while others exhibit higher rates. The potential risk exists for malicious actors to exploit registrars with inconsistent data, allowing them to evade detection systems that rely on the availability and reliability of registration data. An analysis of the extent of malicious domains managed by such inconsistent registrars could offer valuable insights into evasion strategies. 
%Moreover, further research is needed to investigate the extent to which these inconsistencies could affect security systems that depend on registration data.

To facilitate future research, all collected records\footnote{\url{https://doi.org/10.57745/RJX9XH}} and the associated data analysis\footnote{\url{https://github.com/drakkar-lig/whois-right-dataset}} are made public.

\section*{Acknowledgments}
We thank KOR Labs and Sourena Maroofi (KOR Labs) for their valuable support in parsing the registrar's data. 
This work has been partially supported by the French Ministry of Research 
projects PERSYVAL-Lab under contract ANR-11-LABX-0025-01 and DiNS under contract
ANR-19-CE25-0009-01.

\bibliographystyle{splncs04}
\bibliography{whois_vs_rdap, non_academic}

\appendix

\section{Examples of records}\label{app:examples}
\begin{figure*}
    \centering
    \begin{verbatim}
   Domain Name: GOOGLE.COM
   Registry Domain ID: 2138514_DOMAIN_COM-VRSN
   Registrar WHOIS Server: whois.markmonitor.com
   Registrar URL: http://www.markmonitor.com
   Updated Date: 2019-09-09T15:39:04Z
   Creation Date: 1997-09-15T04:00:00Z
   Registry Expiry Date: 2028-09-14T04:00:00Z
   Registrar: MarkMonitor Inc.
   Registrar IANA ID: 292
   Registrar Abuse Contact Email: abusecomplaints@markmonitor.com
   Registrar Abuse Contact Phone: +1.2086851750
   Domain Status: clientDeleteProhibited https://icann.org/epp#clientDeleteProhibited
   Domain Status: clientTransferProhibited https://icann.org/epp#clientTransferProhibited
   Domain Status: clientUpdateProhibited https://icann.org/epp#clientUpdateProhibited
   Domain Status: serverDeleteProhibited https://icann.org/epp#serverDeleteProhibited
   Domain Status: serverTransferProhibited https://icann.org/epp#serverTransferProhibited
   Domain Status: serverUpdateProhibited https://icann.org/epp#serverUpdateProhibited
   Name Server: NS1.GOOGLE.COM
   Name Server: NS2.GOOGLE.COM
   Name Server: NS3.GOOGLE.COM
   Name Server: NS4.GOOGLE.COM
   DNSSEC: unsigned
   URL of the ICANN Whois Inaccuracy Complaint Form: https://www.icann.org/wicf/
    \end{verbatim}
    \caption{Registry WHOIS record of \texttt{google.com} obtained from the VeriSign server}
    \label{fig:whois_example}
\end{figure*}

\begin{figure*}
    \centering
    \begin{verbatim}
 {
  "objectClassName": "domain",
  "ldhName": "GOOGLE.COM",
  "links": [{
      "value": "https://rdap.verisign.com/com/v1/domain/GOOGLE.COM",
      "rel": "self",
      "href": "https://rdap.verisign.com/com/v1/domain/GOOGLE.COM",
      "type": "application/rdap+json"
    },{
      "value": "https://rdap.markmonitor.com/rdap/domain/GOOGLE.COM",
      "rel": "related",
      "href": "https://rdap.markmonitor.com/rdap/domain/GOOGLE.COM",
      "type": "application/rdap+json"}],
  "entities": [{
      "objectClassName": "entity",
      "handle": "292",
      "roles": ["registrar"],
      "publicIds": [{"type": "IANA Registrar ID","identifier": "292"}],
      "vcardArray": [
        "vcard", [
          ["version",{},"text","4.0"],
          ["fn",{},"text","MarkMonitor Inc."]]],
      "entities": [{
          "objectClassName": "entity",
          "roles": ["abuse"],
          "vcardArray": ["vcard",[
              ["version",{},"text","4.0"],
              ["fn",{},"text",""],
              ["tel",{"type": "voice"},"uri","tel:+1.2086851750"],
              ["email",{},"text","abusecomplaints@markmonitor.com"]]]}]}],
  "events": [
    {"eventAction": "registration", "eventDate": "1997-09-15T04:00:00Z"},
    {"eventAction": "expiration", "eventDate": "2028-09-14T04:00:00Z"},
    {"eventAction": "last changed", "eventDate": "2019-09-09T15:39:04Z"},
    {"eventAction": "last update of RDAP database", "eventDate": "2023-05-26T13:57:10Z"}],
  "nameservers": [
    {"objectClassName": "nameserver","ldhName": "NS1.GOOGLE.COM"},
    {"objectClassName": "nameserver","ldhName": "NS2.GOOGLE.COM"},
    {"objectClassName": "nameserver","ldhName": "NS3.GOOGLE.COM"},
    {"objectClassName": "nameserver","ldhName": "NS4.GOOGLE.COM"}],
  }
    \end{verbatim}
    \caption{Part of the Registry RDAP record of \texttt{google.com} obtained from the VeriSign server}
    \label{fig:rdap_example}
\end{figure*}

\end{document}